\documentclass[acmlarge,manuscript,screen]{acmart}

\AtBeginDocument{%
  \providecommand\BibTeX{{%
    \normalfont B\kern-0.5em{\scshape i\kern-0.25em b}\kern-0.8em\TeX}}}

\usepackage{url,hyperref,lineno,microtype,subcaption}

\setcopyright{acmcopyright}
\copyrightyear{2018}
\acmYear{2018}
\acmDOI{XXXXXXX.XXXXXXX}

\acmConference[CHCI '22]{Make sure to enter the correct conference title from your rights confirmation email}{August 18--21,2022}{China}

\begin{document}

\title[EEG in Music Preference]{Introducing EEG Analyses to Help Personal Music Preference Prediction}

\author{Zhiyu He}
\email{hezy22@mails.tsinghua.edu.cn}
\author{Jiayu Li}
\email{jy-li20@mails.tsinghua.edu.cn}
\author{Weizhi Ma}
\email{mawz@tsinghua.edu.cn}
\author{Min Zhang}
\authornote{Min Zhang is the corresponding author}
\email{z-m@tsinghua.edu.cn}
\author{Yiqun Liu}
\email{yiqunliu@tsinghua.edu.cn}
\author{Shaoping Ma}
\email{msp@tsinghua.edu.cn}
\affiliation{%
  \institution{\\Department of Computer Science and Technology, 
Institute for Artificial Intelligence,\\
Beijing National Research Center for Information Science and Technology,
\\Tsinghua University}
  \city{Beijing}
  \country{China}
}

\renewcommand{\shortauthors}{He, et al.}

\begin{abstract}
Nowadays, personalized recommender systems play an increasingly important role in music scenarios in our daily life with the preference prediction ability. 
However, existing methods mainly rely on users' implicit feedback~(e.g., click, dwell time) which ignores the detailed user experience. 
This paper introduces Electroencephalography~(EEG) signals to personal music preferences as a basis for the personalized recommender system. 
To realize collection in daily life, we use a dry-electrodes portable device to collect data. 
We perform a user study where participants listen to music and record preferences and moods. Meanwhile, EEG signals are collected with a portable device. 
Analysis of the collected data indicates a significant relationship between music preference, mood, and EEG signals. 
Furthermore, we conduct experiments to predict personalized music preference with the features of EEG signals. Experiments show significant improvement in rating prediction and preference classification with the help of EEG. Our work demonstrates the possibility of introducing EEG signals in personal music preference with portable devices. Moreover, our approach is not restricted to the music scenario, and the EEG signals as explicit feedback can be used in personalized recommendation tasks. 
\end{abstract}



\maketitle

\section{Introduction}
\label{sec:intro}

In the era of information explosion, the recommender system plays an increasingly important role, which helps users filter preferred items, such as products, news, and music~(\cite{bobadilla2013recommender}).
The most difficult part of a recommender system is accurately predicting users' preferences. However, existing methods mainly rely on users' implicit feedback such as click and dwell time, and attempt to connect them with users' subjective feelings. But this implicit feedback is usually inaccurate and noisy. It ignores the detailed user experience and may be different from the true feelings.  

Recently, with the development of neuroscience technology~(e.g., Electroencephalogram~(EEG)), access to users' explicit feedback to circumvent the problems mentioned above has become possible. Since EEG records the firing information of neuronal activity in the brain, it attracts great attention and is applied to research in text inputting, controlling PC~\cite{liu2020challenges}, and brain activities analysis, especially mood recognition~\cite{zheng2014emotion}. While mood plays an essential role in personal preference judgment, EEG can be decoded as explicit feedback. Thus, we introduce brain signals into the personalized recommender system.

Wet electrodes are commonly used to measure EEG signals. The application of conductive gel on skin yield strong EEG signals. However, these processes are typically troublesome for users. A measurement device that can be realized in people's daily life is particularly important. \citet{Lin2017foreheadEEG} proposes to apply EEG in real-life settings.
Therefore, we use a commercially available EEG Bluetooth headset, a comfortable portable device to measure the users' forehead using dry electrodes. It can provide users with easy and fast daily monitoring to collect the brain signals when they listen to music effectively. 
The recommender system can infer users' moods and experiences with EEG signals, which can be used as explicit real-time feedback.

Nevertheless, inferring personal music preference with portable devices is challenging. Collecting EEG signals with dry electrodes may cause a low signal-to-noise ratio~(\cite{Lin2017foreheadEEG}). As suggested by \citet{cross2001music}, music preference judgment is a higher mental process that involves complex cognitive behaviors. Therefore, a sizeable cognitive gap exists between music preferences and EEG signals collected with portable devices. Considering the above challenges, we pay great attention to data preprocessing in this work.


Since music is closely related to daily life, we conduct an experiment in the music scenario. 
Psychologists have studied the associations between mood and music. \citet{chao2015induced} clarifies the impact of different self-centered moods on music preference, which shows that current mood influences music choice. On the other hand, music is universal partly because it regulates the effect on mood~\cite{swathi2015emotion}. The so-called Music Mood Induction Procedure~(MMIP) relies on music to produce changes in experience and has been utilized to study the impact on moods~\cite{daniel2001emotion}.
Since mood affects the choice of music, it is crucial to get people's emotions in time. However, without interruption, we could not have explicit labels on users' emotions. Our work shows that EEG signals from the portable device reflect moods. From this point, perceived preference is possible. 

Thus, in our work, we collected EEG signals when listening to music and the feedback of preference and mood the participant labeled for every track of music.
Then, we preprocess data for noise reduction and extract typical EEG features based on the collected data. Then we show the relationship between EEG and preference as well as mood.
In the end, we conduct preference prediction with EEG signals. User, music, and EEG features are fused as input for preference prediction. The performance shows a significant improvement by EEG signals.

The main contributions of this work are as follows:

\begin{itemize}
    \item We propose a direction to introduce EEG signals in personal music preference. The performance significantly improved with the help of EEG signals, showing the power of EEG signals to identify users' explicit feedback. 
    \item  A user study is conducted in music scenarios. We preprocess EEG signals and analysis the preference association. We discuss the influence of EEG features over electrodes and frequency bands. 
    \item We collect the data by portable EEG device, which reveal a new way for future new generation music recommendation when portable devices are popular and convenient~\footnote{The code and data is available at https://github.com/hezy18/EEG\_music\label{foot:link}}. 
\end{itemize}

The remainder of this paper is organized as follows: We review related work in Section~\ref{sec:related}. Then we introduce our user study methodology in Section~\ref{sec:lab_study} and provide analysis and findings in Section~\ref{sec:analysis}.
Section~\ref{sec:Predction} presents the experiments for preference prediction and further analysis.
And we discuss our concerns and limitations in Section~\ref{sec:discussion}. Finally, Section~\ref{sec:conclusion} provides the conclusions of our work.

\section{Related Work}
\label{sec:related}
In this section, we review related work on personalized music recommender systems, EEG-based preference detection, and other EEG-related research. 

\subsection{Personalized Music Recommender Systems}

Music is one of the most critical scenarios for recommender systems.

There are many proposals for recommendation methods, most of which are based on collaborative filtering~(CF)~\cite{wu2021survey}. 
These approaches made predictions about a user's preference~(e.g., rating) on an item based on the users or items with similar preferences. 
Collaborative filtering based on user id and item id has also been applied to music recommendation successfully~\cite{song2012survey}.

Since users' music preferences are dynamic and diverse, music recommendation relies on context attributes~\cite{deldjoo2021content,murthy2018content}. In conventional content-based music recommendation, music audio signals and descriptive metadata were used as item-level content~\cite{hansen2020contextual,yoshii2006hybrid,van2013deep,sachdeva2018attentive,banff2004content}. 
In Cheng's work~\cite{cheng2016effective}, the characteristics of venues and music were mapped into a latent semantic space to measure the similarity between music and locations. 

As more user profiles can be collected from online behaviors, user-generated context information has been considered to improve the recommendation performance. \citet{yapriady2005combining} applied user demographic data, and \citet{shen2020peia} extracted emotion-oriented user features from the user blog text for music recommendations.
The social context was also considered, such as online social networks on Facebook~\cite{mesnage2011music}, social influence~\cite{chen2019improving}, and cultures of users~\cite{zangerle2018culture,schedl2020listener}.

However, these works focused on the context data on the Internet, which is not real-time.
With real-time EEG signals in the physical world, it is possible to predict users' music preferences in real-time. 
Moreover, a better understanding of users' preferences will lead to better recommendation results. Traditional music recommender systems mainly use implicit feedback from users, such as click and dwell time. 
But hearing does not mean preferring, and the clicks do not show preference. As for explicit feedback such as rating, it is difficult for users to provide it every time proactively. Instead, we perform personalized recommendations with EEG signals from the portable device to capture the real-time changes in users' music preferences. It can replace explicit feedback, which is hard to obtain.

\subsection{EEG Based Preference Detection}

Several research works have been proposed to use EEG signals in predicting the self-reported preferences~(in terms of ratings) and the consumers' behavior~(e.g., the choice of the products) in various scenarios.

In some works, a recommendation framework is proposed with the use of EEG to obtain the psychological state before information need is expressed. 
\citet{Yadava2013ecommerce} proposed a framework of a recommendation system for e-commerce products by fusing the pre-purchase and post-purchase ratings. EEG signals of users have been recorded while watching 3D products. The relative power of EEG was used to reflect the pre-purchase rating. 
\citet{Lee2020theraypy} focused on music scenarios to provide users for depression therapy with a list of soothing music tracks. 
These previous works proposed the concept of replacing part of the preference system with EEG but lacked the analysis and computing of the relationship between EEG and preference. 



\citet{Guibing2013Recommender} proposed a neuro-signal-based framework to predict the consumer choices for online products. 
Participants chose one of three images of the same product type and used EEG signals to predict which product was preferred. 
Features were modeled with HMM classifier and achieved an accuracy of 70.33\%. 
However, it manually restricted the scope of product selection, and the experimental settings differed from the product selection process in practice. 
\citet{kumar2019fusion} proposed a multi-modal framework to integrate physiological signals with product reviews to obtain the overall evaluation of a product. 
It focused on the quality of products rather than a personalized preference for the items. 
Moreover, these works used wet electrode EEG, which sacrifices convenience and is impossible for day-to-day usage.


These previous works indicate that EEG has great potential to capture the preferences of people. 
However, they paid little attention to applying EEG for preference prediction in the personalized recommendation scenario. 
In our work, we try to incorporate the EEG signals explicitly into personalized recommendations. 
Furthermore, 
dry electrode portable device could be used in daily life in the future. 
So we implement our experiments with commercial dry electrode devices for EEG collection. 

Unlike previous work, we genuinely introduce EEG analysis into personalized preference detection in the music recommendation scenario.

\subsection{Other EEG Related Research} 
A growing number of studies are applying EEG signals, such as controlling keyboards and mice, inputting speech and text with thoughts, and monitoring human emotions.

EEG signals have been applied to perform some simple interactions for people in the human-computer-interaction applications. 
\citet{Liu2019BCIvisual} tried to develop a BMI system composed of a spelling module and a web-browsing tool. They trained an SVM model to predict the target button based on EEG signals and showed that participants could use some basic functions of a web search engine with the system. 
\citet{Perego2011cognitive} designed a hand-free BCI-based search system that illustrates the possibility of using BCI to replace keyboard and mouse in real life. 
Recently, EEG's potential in information retrieval has been proposed by \cite{liu2020challenges}.
\citet{chen2021search} proposed a prototype for virtual typing and searching tasks on computers with EEG devices.  \citet{Nuyu2018cortical} showed that people with tetraplegia could use an intracortical BMI to control a computer to use some popular applications, e.g., web browsing, email, and chatting.

Furthermore, EEG signals have made inputting text and speech with thoughts possible. 
\citet{christian2015text} showed for the first time that continuously spoken speech can be decoded into the expressed words from BMI recordings. Their system can achieve word error rates as low as 25\% and phone error rates below 50\%. 
Furthermore, brain signals could be translated into speech based on a neural decoder that explicitly leverages kinematic and sound representations encoded in human cortical
activity to synthesize audible speech~\cite{anuman2019speech}. 

In the psychological scenario, EEG signals have been utilized to recognize and analyze human moods. 
In Zheng's work~\cite{zheng2014emotion}, the combination of deep belief networks and the hidden Markov model was used to classify two kinds of emotions~(positive and negative) using EEG signals. 
\citet{soleymani2012multimodal} proposed a user-independent emotion recognition method to recover effective tags for videos using EEG signals, pupillary response, and gaze distance.

These existing research efforts show the application potential of EEG in various fields. 
We believe EEG is also helpful for personalized recommendation scenarios.
However, existing research is not feasible to apply to everyday life as they are mainly based on the wet-electrode EEG device or invasive electrodes. 
In our work, we use a portable dry-electrode device, which has little effect on user experience. 
Although it is a great challenge that the collected signals are with low spatial resolution and poor signal-to-noise ratio, we have seen the prospect of its application~\cite{Lin2017foreheadEEG}.

Besides, previous works show that EEG has a relationship with mood, and music is also related to mood~\cite{chao2015induced,swathi2015emotion}. So we believe mood is also a variable to help personalized music recommendations by indicating users' preferences.


\section{User Study Methodology}
\label{sec:lab_study}
We conducted a user study to collect EEG signals, moods, and preferences while listening to music. 
Participants were required to wear a commercial EEG Bluetooth headset with dry electrodes, listen to designated music, and give preference and mood feedback. They would be paid according to the experiment time. The whole user study has been reviewed and approved by the Department of Psychology Ethics Committee, Tsinghua University (THU202118). 
This section will introduce our user study procedures and data preprocessing methods. 

\subsection{User Study Procedure}
\subsubsection{Procedure Overview}

\begin{figure*}[hb]
    \centering
    \includegraphics[width=15cm]{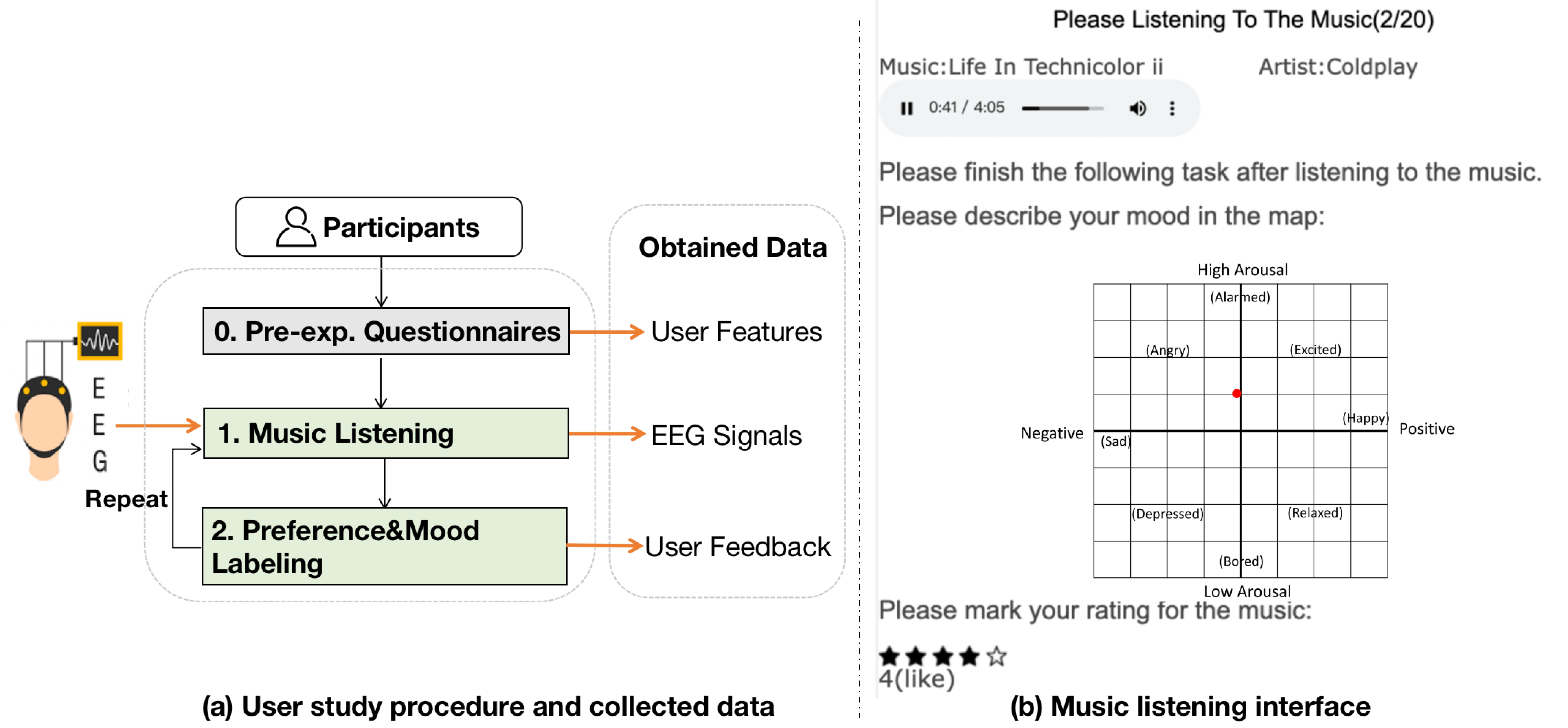}
    \caption{User study procedure and the user-computer interface of the experiments.}
    \label{fig:interface}
\end{figure*}

The overall illustration of the user study procedure is shown in Figure~\ref{fig:interface}(a).

At the beginning of the user study, every participant was well informed of the experiment process and required to sign the informed consent. 
Then questionnaires about their basic information, such as gender and age, would be filled out. 
Then the experiment for music preference and EEG signals collection was conducted. 
Firstly, we helped the participant wear a 6-electrode dry electrode EEG headset and confirmed the Bluetooth connection to ensure the recording and transferring of EEG signals.
Then, Each participant listened to 20 randomly selected tracks. 
They are suggested to listen to each track for at least one minute without moving their body to minimize noise. Then the participant would label their preference rating~(in five levels) and mood~(with a two-dimensional Thayer model map~\cite{thayer1990biopsychology}) in each track. 

\begin{figure}
    \centering
    \includegraphics[width=8cm]{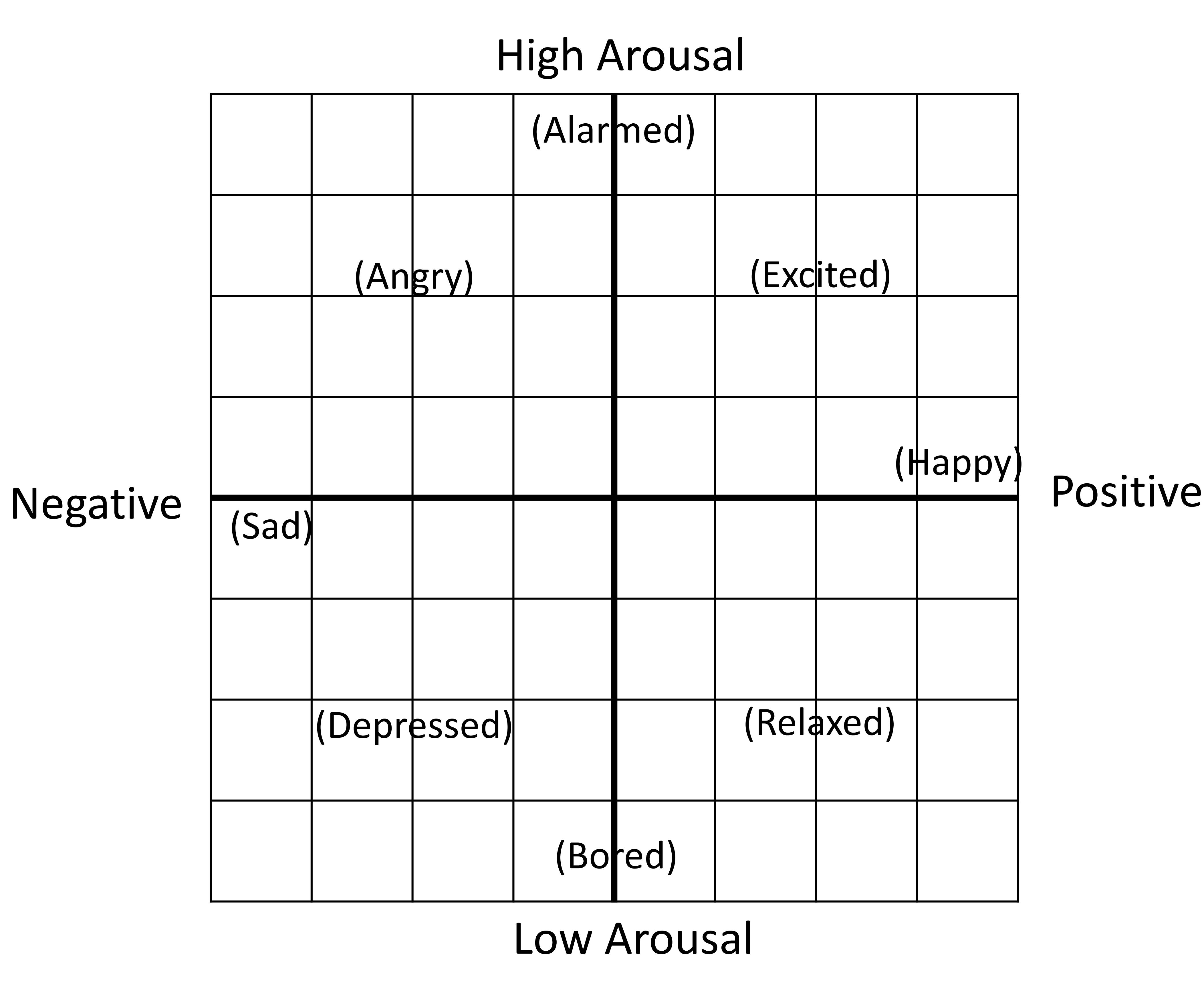}
    \caption{Two-dimensional description map of Thayer mood, where the X-axis represents valence, and the Y-axis represents arousal. }
    \label{fig:thayer_mood}
\end{figure}

A concise website interface was designed for music listening as well as preference and mood labeling. The main page is shown in Figure~\ref{fig:interface}(b). 
As we required participants to record mood, an intuitive method to label mood was necessary. Balancing the label precision and complexity, we implemented the Thayer mood model~\cite{thayer1990biopsychology} with a two-dimensional description map of mood, i.e., valence~(positive or negative) and arousal~(energetic degree), as shown in Figure~\ref{fig:thayer_mood}. During the user study, participants will choose the corresponding location on the page for the mood brought about by listening to the music. Before the experiment, all participants were trained to use the map to describe moods.

After the experiment, we collected the basic information through questionnaires, the EEG signals when listening to music, and the user feedback in terms of preference and mood of each track. The obtained dataset will then be preprocessed.

Participants, music preparation, and apparatus in the experiment are detailed below.

\subsubsection{Participants}

enrolled in this study, and one of them had no data records due to equipment reasons. 
Among the remaining 19 participants, 11 were female and 8 were male, aged 20-26~(M=22.26, SD=1.56\footnote{M for mean value, SD for standard deviation}). 
All participants are native Chinese speakers mastering college-level Chinese reading and writing skills. And they admit that they are right-handed and do not suffer from any neurological disease. 
Signed consent was obtained from each participant before the experiment was carried out. 
It takes about one hour to complete the whole task for each participant, and each participant is paid \$7.91 after they complete the tasks.

\subsubsection{Music data preparation}

We selected 1000 music tracks from the Million Song Dataset\cite{}, which includes audio features and metadata for a million contemporary popular music tracks. 
Then, we divided the \textit{music valence} feature into ten equal intervals and selected the top 1000 popular music tracks from each interval. After music tracks with \textit{demo} or \textit{live} in their names being excluded, 9704 music tracks were left. 
Then, we randomly selected 20 tracks from the remaining 9704 tracks, and all participants were required to listen to these 20 tracks.

\subsubsection{Apparatus} 

\begin{figure*}[ht]
    \centering
    \includegraphics[width=15cm]{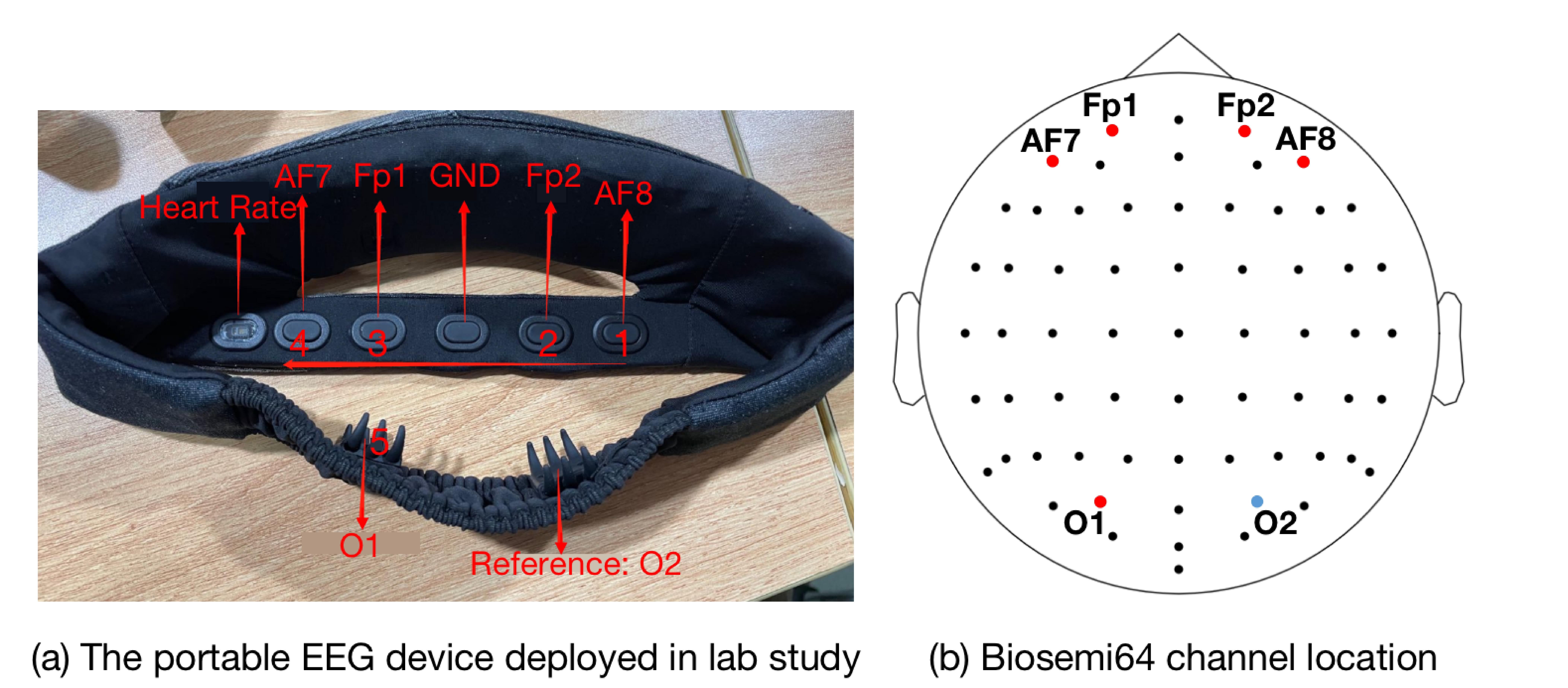}
    \caption{The portable EEG headset deployed in our lab study with the mark of electrodes and their corresponding sensor locations. Red circles are the recorded electrodes, and the blue circle is the reference electrode.}
    \label{fig:device}
\end{figure*}

Our study used a laptop computer with a 17-inch monitor with a resolution of 1,600×900. 
We deployed a 6-dry-electrode headset to capture the participants’ EEG data during the experiment continuously. The electrodes were placed on the scalp, and Figure~\ref{fig:device} shows the 6-electrode EEG cap layout used in this study. The signals were digitized at 250 Hz and stored in a PC for offline analysis. 

Among six electrodes, O2 is the reference electrode. The value of each electrode is represented by the difference between it and O2. Electrode AF8, Fp2, Fp1, and AF7 are in the frontal lobe. O1 and O2 are at occipital lobel.

\subsection{User Behavior and EEG Data Preprocessing}

We achieved a series of data by user study, including user demographics, preferences, moods, and EEG signals. However, we cannot use EEG signals in the data analyses. Alignment, normalization, filtering, and feature extraction need to be fulfilled in this process.

\subsubsection{Data Alignment and Denoising} 
In the user study, each pair of labels~(preferences and moods) was expected to correspond to a segment of EEG signals. 
We regarded the time of clicking on 'music start' as when the EEG signals start. As participants should listen for at least 90 seconds, we truncated each EEG segment for 90s~(that is, 3000 sampling points). We noted the interval time from the 'music start' button to the 'music pause' button of the same piece of music. Cases with less than 60s~(due to the equipment reason) would be removed. Then, For the same electrodes for each participant, we did a min-max normalization to minimize the difference between each participant.

\begin{figure*}[ht]
    \centering
    \includegraphics[width=15cm]{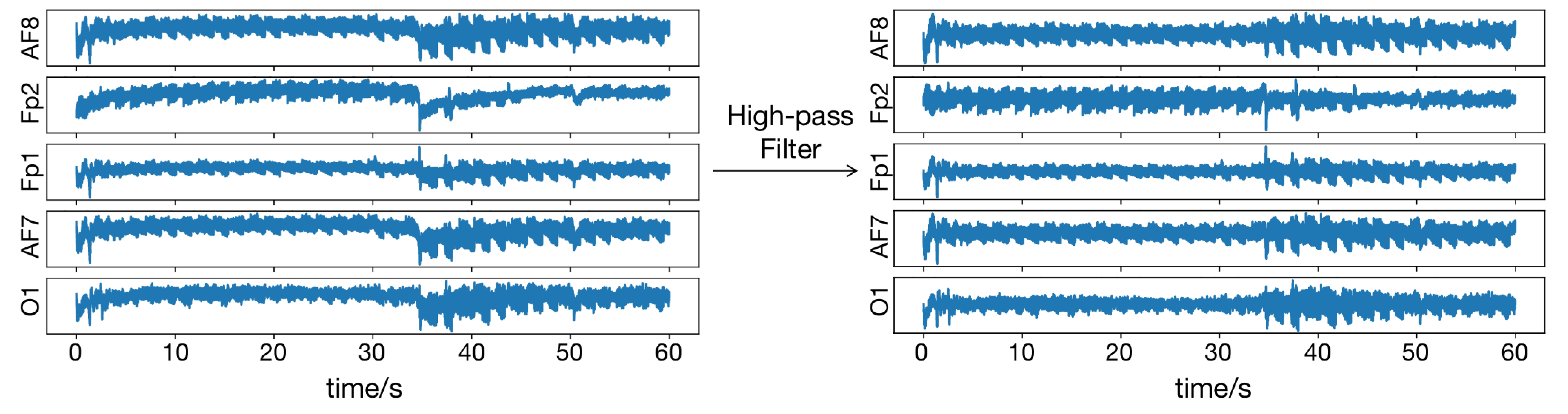}
    \caption{Time series data before and after removing low-frequency drift for a sample.}
    \label{fig:filter}
\end{figure*}

A critical step is filtering the data to remove low-frequency drifts. The slow drifts are problematic because they reduce the independence of the assumed-to-be-independent sources (e.g., during a slow upward drift, the neural activity, heartbeat, blink, and other muscular sources will all tend to have higher values), making it harder for the algorithm to find an accurate solution. Thus, we use a high-pass filter with a 1 Hz cutoff frequency. We enumerate a sample in Figure~\ref {fig:filter}, reflecting the change of time series data before and after filtering.

\subsubsection{EEG Feature Extraction}

While EEG signals look like a waveform that vibrates in a very complex pattern, power spectrum analysis that classifies according to frequency is used frequently when observing EEG. Power spectrum analysis assumes that the EEG signal is a linear combination of simple vibrations that vibrate at a specific frequency. Thus, we can decompose each frequency component in this signal to indicate its magnitude~(or power). In our study, the frequency power of trials was extracted with Welch’s method with windows of 250 samples. The changes of power were averaged over the frequency bands of theta (4-8 Hz), alpha (8-12 Hz), beta (12-30 Hz), gamma (30-45 Hz), and the whole frequency. 

We compute PSD~\cite{GramfortEtAl2013a} over the five frequency band mentioned above of time series at each electrode. Thus, five passbands~(4-8 Hz, 8-12 Hz, 12-30 Hz, 30-45 Hz, and all-pass bands) are combined with five electrodes~(AF8, Fp2, Fp1, AF7, and O1) in pairs. We get a total of 25 EEG features for each case.


\subsubsection{Label Statistics} 
After this process, 319 cases from 19 participants remain. Each case shows a participant's feedback to an item~(i.e., music) with preference and mood labels, as well as a segment of EEG signals during listening to the music. As for the preference~(in terms of rating), cases were rated 1-5. Specifically, among the 319 cases, 35 cases~(10.97\%) were rated 1, 62 cases~(19.44\%) were rated 2, 112~(35.11\%) were rated 3, 77~(24.14\%) and 33 cases~(10.34\%) were rated 4 and 5 respectively. Most records were rated 2, 3, and 4. 

\begin{figure*}[ht]
    \centering
    \includegraphics[width=15cm]{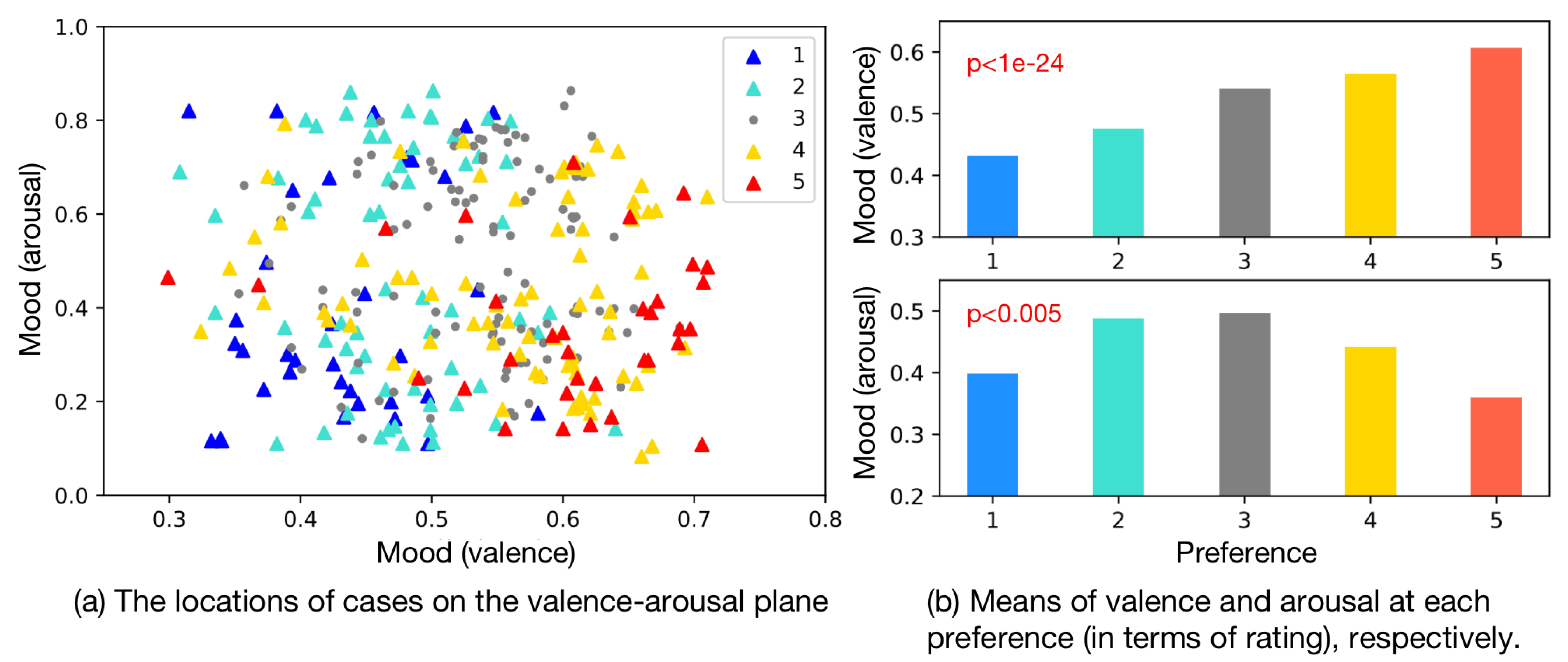}
    \caption{The distribution of mood~(valence and arousal) in each preference~(in terms of rating) and the relationship between mood and preference. The colors represent preferences(in terms of ratings). P stands for $p$ value from the ANOVA test between 5 groups(rating 1-5).}
    \label{fig:mood_pref}
\end{figure*}

Next, we analyse the relationship between mood and preference. Figure~\ref{fig:mood_pref}(a) shows the locations of each cases on the arousal-valence plane with the colors standing for preference. Valence ranged 0.299-0.71~(M=0.528, SD=0.093) and arousal ranged 0.083-0.863~(M=0.456, SD=0.212). 
The magnitude of arousal is greater than that of valence. 
Furthermore, preferences tend to be positively correlated to valences, and extreme preferences~(i.e., rating 1 and 5) are usually accompanied with low arousal.
Figure~\ref{fig:mood_pref}(b) better indicates this conclusion. 
The mean values in each preference category demonstrate that as the preference increases, the valence also increases~(F-statics=34, p-value\textless1e-23~\footnote{Results come from ANOVA test}). 
Meanwhile, moderate preference corresponds to higher arousal, and extreme preference tends to have lower arousal~(F-statics=4, p-value\textless0.005). The analysis shows that different preferences bring different moods. Moreover, valence and arousal have different meanings in the description of preference.

\section{EEG Analysis and Findings}
\label{sec:analysis}

We found a significant relationship between moods and preference in the previous section. 
In this section, we provide a statistical analysis of music preferences, mood records, and EEG signals in the user study to explore the usefulness of EEG features. 

\subsection{Relationship Between Music Preference and EEG signals} 

\begin{figure*}[ht]
    \centering
    \includegraphics[width=15cm]{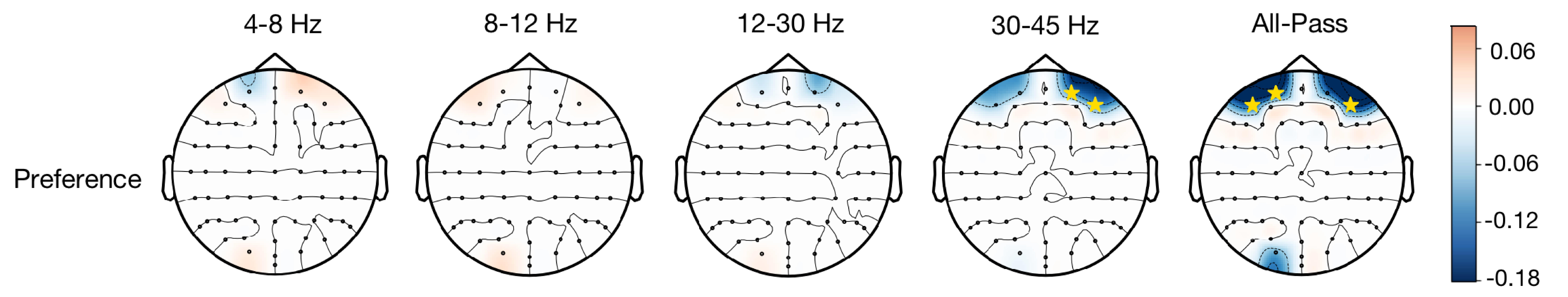}
\caption{The correlations (overall cases) of the preference~(in terms of rating) with PSD in the broad frequency bands of theta (4-8 Hz), alpha (8-12 Hz), beta (12-30 Hz), gamma (30-45 Hz) and all-pass. The highlighted stars indicate a significant correlation~(p\textless0.05) between preference and PSD features.}
    \label{fig:pref_corr}
\end{figure*}

To explore the correlation between EEG and preference, 
we compute the Pearson correlated coefficients between the PSD and preference~(in terms of rating, in the following, we will mix the usage of terms \textit{preference} and \textit{rating}) of each case. Figure~\ref{fig:pref_corr} shows the correlations with significantly~(p\textless0.05) correlating electrodes highlighted. For electrodes with no records, we set the correlation coefficient to 0. 
We can find that as the frequency of the passband increases, the overall correlation of PSD and preference turns positive to negative. 
The correlation coefficient ranged from -0.18 to 0.03, demonstrating that the EEG signals vary greatly at different frequencies in different locations.  

Generally, the PSD and preference have a certain linear correlation at the partial electrode of the partial band. 
The PSD at electrodes AF8 and Fp2 in the broad frequency bands of 30-45 Hz significantly correlate. Besides, PSD at all electrodes except O1 in the all-pass band has a relatively higher correlation with preference, which is a negative correlation. Therefore, we further analyze these specific electrodes for 30-45 Hz and all-pass band. 

\begin{figure*}[ht]
    \centering
    \includegraphics[width=15cm]{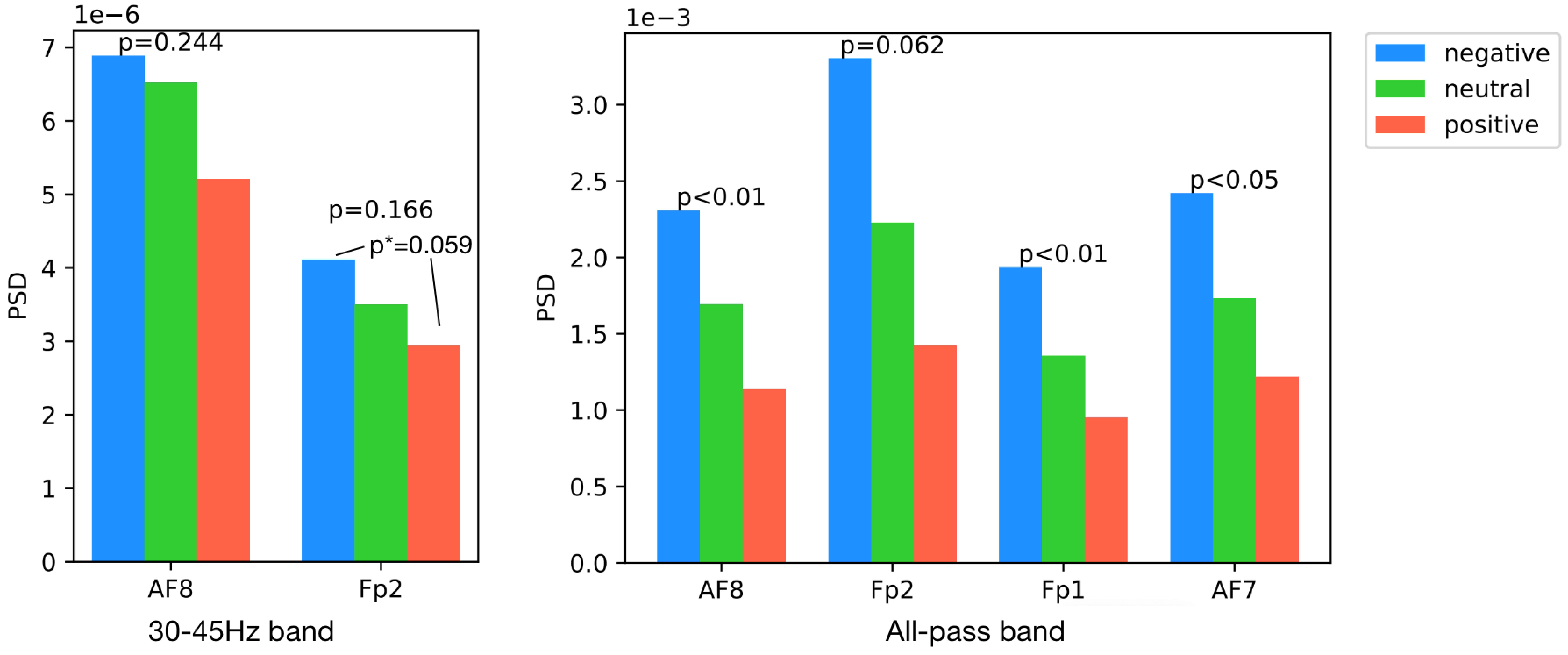}
    \caption{Mean values of PSD at specific EEG electrodes for 30-45Hz band and all-passed band in negative, neutral, and positive preference. $p$ indicates P values from the ANOVA test between 3 groups of preference. $p^{*}$ stands for p-value of a two-sided t-test conducted between negative and positive groups for 30-45 Hz band.}
    \label{fig:pref_mean}
\end{figure*}

In the self-assessment of preferences, a closed question~(like it or not) is easier to answer than an open one~(the degree of preference). 
Thus,
we divide the cases into three groups according to the scores of preference~(ranged from 1 to 5, discretely). Scores 1 and 2 are considered to be negative, label 3 is represented by neutral, and 4-5 are positive. The number of cases in each group is 97, 112, and 110, which is almost evenly distributed. 

For each specific EEG electrode, we compute the mean values of PSD for the all-pass band and 30-45Hz band. Then we conduct an analysis of variance between PSD in three groups. Figure~\ref{fig:pref_mean} shows the Analysis results.

As for AF8 at 30-45 Hz band, The p-value from Pearson correlation in Figure~\ref{fig:mood_corr} is 0.048, almost near to 0.05. Besides, we further analyze the standard deviation of each group. Compared with Fp2 at 30-45 Hz band, groups of PSD in AF8 have greater deviation, which brings more confounding and insignificant factors. 
As the preference changes from positive to negative, the PSD value decreases overall. 
Besides, as the neutral group is more like in the fuzzy zone, positive and negative preference is easier to distinguish. 
The difference between the three groups~(positive, negative, and neutral) is significant, especially at electrodes AF8, Fp1, and Af7 for the all-pass band, which demonstrates the significance and capabilities of the EEG signals at the all-pass band.

\subsection{Relationship Between Mood and EEG signals} 

\begin{figure*}[hb]
    \centering
    \includegraphics[width=15cm]{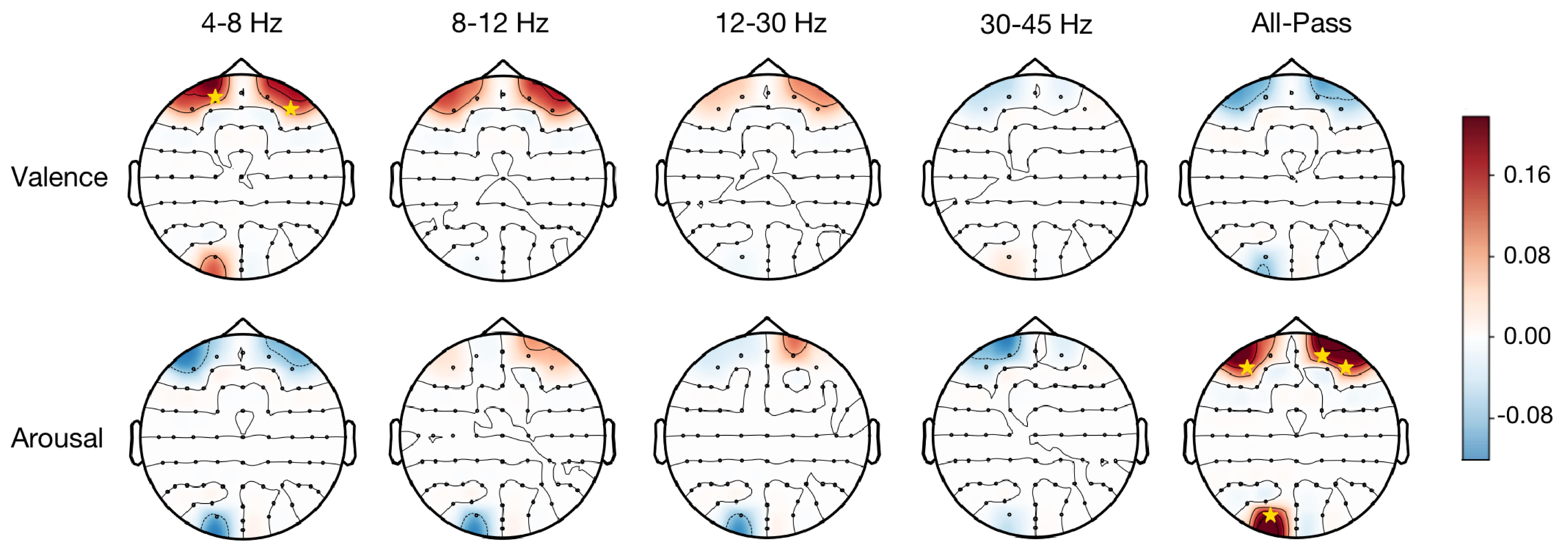}
    \caption{The mean correlations (overall participants) of the valence and arousal with PSD in the broad frequency bands of theta (4-8 Hz), alpha (8-12 Hz), beta (12-30 Hz), gamma (30-45 Hz) and all-pass. The highlighted stars correlate significantly (p\textless0.05) with the valence or arousal.}
    \label{fig:mood_corr}
\end{figure*}

Correlation statistics are also applied to the relationship between mood and EEG signals. The heat in Figure~\ref{fig:mood_corr} represents the coefficient computed by Pearson correlation between each case's power changes and preference. Significant electrodes~(p\textless0.05) are highlighted. 
As the frequency of the passband changes from low to high, the correlation with valence changes from positive to negative, while the correlation with arousal changes from negative to positive. Thus, the mood is depicted in two dimensions that correlate differently with EEG signals. 
Besides, PSD at electrodes for the 4-8 Hz band has a relatively higher correlation with valence, which is a positive correlation. Meanwhile, PSD at electrodes for the all-pass band has a significant positive correlation with arousal. 
Therefore, we further analyze the 4-8 Hz band for the valence dimension and the all-pass band for the arousal dimension. 

As for mood, we divide the cases into three groups according to valence and arousal~(ranged 0-1, continuously), respectively. 
On a 0-1 scale, the threshold for dividing positive and negative is 0.5, which is a clear line when self-assessing valence. Besides, 35\% of cases are negative. People are more inclined to label positive moods. Moods without clear positive and negative tendencies are more likely to be marked in the positive position in the middle. Thus, we use 0.5 and 0.6 as the threshold to divide negative, neutral, and positive groups. As for arousal, we use 0.33 and 0.67 ~(the trisection point between 0-1) as the threshold to divide the three groups. At last, the number of cases in each group is 107, 136, 76, and 107, 136, 76 for valence and arousal, respectively. 

\begin{figure*}[ht]
    \centering
    \includegraphics[width=15cm]{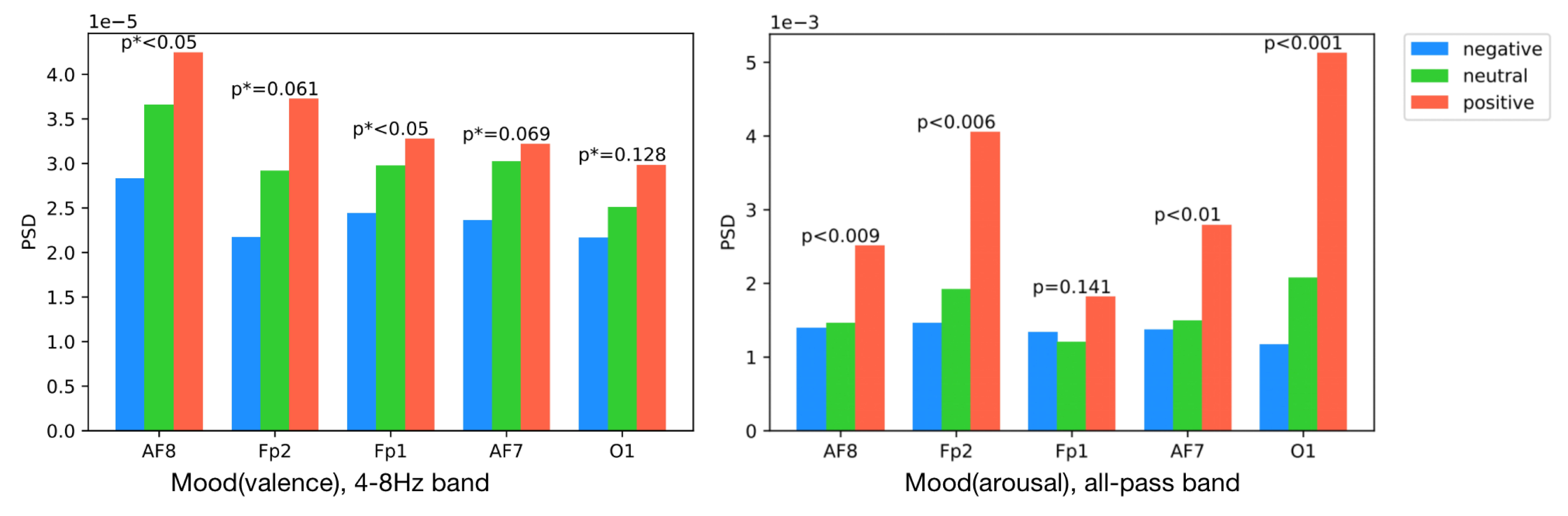}
    \caption{Mean values of PSD on 4-8 Hz band and all-pass band~(in terms of different EEG electrodes) in negative, neutral, and positive groups divided by valence~(left) and arousal~(right) scores. \textbf{p} stands for the p values from the ANOVA test between negative, neutral, and positive groups.}
    \label{fig:mood_mean}
\end{figure*}

Also, for each EEG electrode, we compute the means of PSD at all electrodes for the 4-8 Hz and all-pass bands. Then we analyze the variance between PSD in three groups for valence and arousal, respectively. 
As shown in Figure~\ref{fig:mood_mean}~(left), 
greater valence gets higher EEG power, while there is little difference between neutral and positive. At electrode Fp1, The difference between negative, neutral, and positive groups is significant. 
As for arousal, we can see the obvious distinction between the three groups in Figure~\ref{fig:mood_mean}~(right). The difference is significant at electrodes AF8, Fp2, AF7, and O1. High arousal has high EEG power overall, especially in the positive group. 

As the arousal dimension is more direct, we see a stronger relationship with arousal than with valence using EEG. The relationship between mood and EEG is closer than the relationship with preference, which is also in line with our perception. As mood and preference are related, adding nonlinear factors, EEG can better describe preference.  

\begin{table}[h]
\renewcommand\arraystretch{1.5}
\setlength{\abovecaptionskip}{0.2cm}
\setlength{\belowcaptionskip}{0.0cm}
\begin{tabular}{@{}l|ccccc|ccccc@{}}
\hline
 & \multicolumn {5} {c|} {\textbf{Valence, 4-8 Hz band}} & \multicolumn {5} {c}\ {\textbf{Arousal, all-pass band}}\\
 & AF8  & Fp2 & Fp1 & AF7 & O1 &  AF8 & Fp2 & Fp1 & AF7 & O1 \\ 
\hline
Negative \& Neutral  & 2.33$^{*}$  & 1.63  & 2.89$^{*}$  & 2.39$^{*}$  & 1.33  & -0.20 & -0.77 & 0.49 & -0.35 & -1.24\\
Neutral \& Positive & 0.09  & 0.56  & 0.62  & 0.53  & 0.36  & 2.60${^*}$ & 2.31${^*}$ & 1.96 & 2.47${^*}$ & 2.57$^{*}$\\
Negative \& Positive & 1.95  & 2.09$^{*}$  & 2.02$^{*}$  & 1.58  & 1.59  & 2.68$^{**}$ & 2.91$^{**}$ & 1.40 & 2.73$^{**}$ & 3.50$^{**}$  \\
\hline
\end{tabular}
\caption{T statistics by two-side t-test of PSD on 4-8 Hz band and all-pass band~(in terms of different EEG electrodes) in positive, neutral, and negative groups divided by valence and arousal scores, respectively. $^{*}$and ${^**}$ indicate p-value\textless0.05 and p-value\textless0.01, respectively.}
\label{tab:mood_ttest}
\end{table}

We further analyze the difference between the two groups. A two-side independent t-test is conducted between negative and neutral groups, neutral and positive groups, and negative and positive groups. The result is shown in Table~\ref{tab:mood_ttest}. 
As for valence, the difference between the negative and neutral groups is more significant than the difference between the neutral and positive groups is significantly different from the other two groups. Thus, the negative group is more separable. 
As for arousal, the positive group is more different from the other groups. 
Thus, negative valence and high arousal have better differentiation compared with other conditions using PSD features of EEG signals. 


\section{EEG-aware Music Preference Prediction Experiments}
\label{sec:Predction}

Since we have found a close relationship between EEG signals and music preference, EEG signals and mood, as well as mood and preference, we further introduce EEG signals to predict music preferences~(in terms of ratings)~\footnote{The code implementation is available at https://github.com/hezy18/EEG\_music}. 
We use representative models for rating prediction and preference classification tasks to explore the improvement by adding EEG. 

\subsection{Experiment Settings}
In this section, we will introduce the basis experiment settings of our work. 

\subsubsection{Baseline Features} 
Considering that traditional recommender systems predict preference by the user and item features, we use these features as the baseline. Age and gender represent each \textbf{user}, and \textbf{music} is described with 12 audio features from Spotify\footnote{https://developer.spotify.com/console/get-audio-features-track/}. 
Detailed features are shown in Table~\ref{tab:baseline_features}. 

It's worth mentioning that we embed the feature name "tags" from Spotify. We keep the top 14 most common labels and convert them to a 0-1 sequence using a multi-label binarizer. 

\begin{table}[hb]
\renewcommand\arraystretch{1.5}
    \centering
    \scalebox{0.9}{
    \begin{tabular}{c|ccc|ccc}
    \hline
    \textbf{Group} & \textbf{Source} & \textbf{Feature Name} & \textbf{Data type} & \textbf{Source} & \textbf{Feature Name} & \textbf{Data type} \\
    \hline
    user & Questionnaire & age & int & Questionnaire & gender & bool\\
    \hline
    & Spotify & popularity & int & Spotify & year & int \\
    & Spotify & danceability & float  &Spotify & energy &  float \\
    item & Spotify & loudness & float & Spotify & speechiness & float \\
    & Spotify & valence & float &  Spotify &  acousticness & float\\
    & Spotify & instrumentalness & float & Spotify & liveness & float\\
    & Spotify & tempo & float & Spotify & tags & 0-1 sequence \\
    \hline
    \end{tabular}
    }
    \caption{Fourteen features for baseline were extracted from the questionnaire and Spotify acoustic feature set.}
    \label{tab:baseline_features}
\end{table}

\subsubsection{Based models} 
To explore the effect of adding EEG on a variety of typical models, we choose based models from both traditional machine learning models and neural models. The selected model and related settings are described below. 


As for traditional machine learning models, we use models with ensemble learning.
\textbf{Gradient Boostintg Decision Tree~(GBDT)}, \textbf{Random Forest~(RF)}\cite{scikit-learn} and eXtreme Gradient Boosting~\cite{xgboost}(XGBoost).
\textbf{Multi-layer Perceptron~(MLP)} is used as neural model. We conduct three layers. LBFGS was used to optimize the loss of squared error and log-loss function for rating prediction and preference classification, respectively.

\subsection{Rating Prediction}

Recommender system is based on predicting the user's preference for the item. To introduce EEG into recommendation, we start with EEG prediction preference~(in terms of rating, ranging from 1 to 5). Five times 5-fold validation are conducted on both baseline and PSD-added condition. The task is music and participant dependent. The performance is estimated by Mean Squared Error~(MSE) and is shown in Table\ref{tab:regression} . We fairly call out the best results for each model under various parameters.

It can be seen that after adding EEG, the effect of prediction is improved significantly on these typical models. XGBoost has improved for 23.5\% from the baseline. The best results are obtained with MLP with the mean MSE over 5 tiems of 5-fold cross validation is 0.9025.

\begin{table}[ht]
\renewcommand\arraystretch{1.5}
\centering
\setlength{\abovecaptionskip}{0.5cm}
\setlength{\belowcaptionskip}{0.5cm}
\begin{tabular}{@{}c|cccc@{}}
\hline
\textbf{Features}& \textbf{GBDT} & \textbf{RF} & \textbf{XGBoost} & \textbf{MLP} \\ 
 \hline
user,item  & 0.9801 & 0.9811 & 1.1953 & 0.9471\\
 \hline
user,item,PSD  & 0.9274$^{**}$ & 0.9336$^{**}$ & 0.9146$^{**}$ & \underline{0.9025}$^{*}$\\
 & (-5.4\%) & (-4.8\%) & (-23.5\%) & (-4.7\%)\\
\hline
\end{tabular}
\caption{Overall performance of baseline method~(using users and items features) and our method~(adding EEG signals, specifically, the PSD) to predict ratings in terms of Mean Squared Error~(MSE, the lower the better). Two-side t-test is conducted for 5 times 5-fold cross validation. $^{*}$and $^{**}$ indicate p-value\textless0.05 and p-value\textless0.005, respectively. underline{Underline} indicates the best results.}
\label{tab:regression}
\end{table}

\subsection{Preference Classification}

In the self-assessment of preferences, a closed question~(like it or not) is easier to answer than an open one~(the degree of preference). Therefore, in practical applications, it is sufficient for people to use binary labels of preferences. 
Thus, we conduct the task to classify preference into three groups~(negative, neutral and positive) and use accuracy as evaluation metric. Also, five times 5-fold validation are conducted with music and participant dependent for both baseline and PSD-added condition. We call out the best performance with parameter tuning for each model. 

\begin{table}[h]
\renewcommand\arraystretch{1.5}
\centering
\setlength{\abovecaptionskip}{0.0cm}
\setlength{\belowcaptionskip}{0.5cm}
\begin{tabular}{@{}c|cccc@{}}
\hline
\textbf{Features} &  \textbf{GBDT} & \textbf{RF} & \textbf{XGBoost} & \textbf{MLP} \\ 
\hline
user,item  & 0.4540 & 0.4351 & 0.4150 & 0.4639 \\
\hline
user,item,PSD & 0.4703 & 0.4915$^{**}$ & 0.4865$^{*}$ & \underline{0.4927}$^{*}$\\
 & (+3.6\%) & (+13.0\%) & (+5.4\%) & (+6.2\%) \\
\hline
\end{tabular}
\caption{Overall performance of baseline method~(using users and items features) and our method~(adding EEG signals, specifically, the PSD) to classify preference into negative, neutral and positive groups. The performance is estimated by accuracy~(the higher the better). Two-side t-test is conducted for 5 times 5-fold cross validation. $^{*}$and $^{**}$ indicate p-value\textless0.05 and p-value\textless0.005, respectively. underline{Underline} indicates the best results.}
\label{tab:performance}
\end{table}

The performance of models improve with the help of EEG. Random forest~(RF) improved 13.0\% from the baseline. The best results are obtained with MLP with the mean accuracy over 5 times of 5-fold cross validation is 0.4927. 

\subsection{Feature Analysis}
To further analyse the importance of difference features how they affect the model, we conduct ablation study and feature importance analysis.

\subsubsection{Ablation Study}
\begin{figure*}[ht]
    \centering
    \includegraphics[width=15cm]{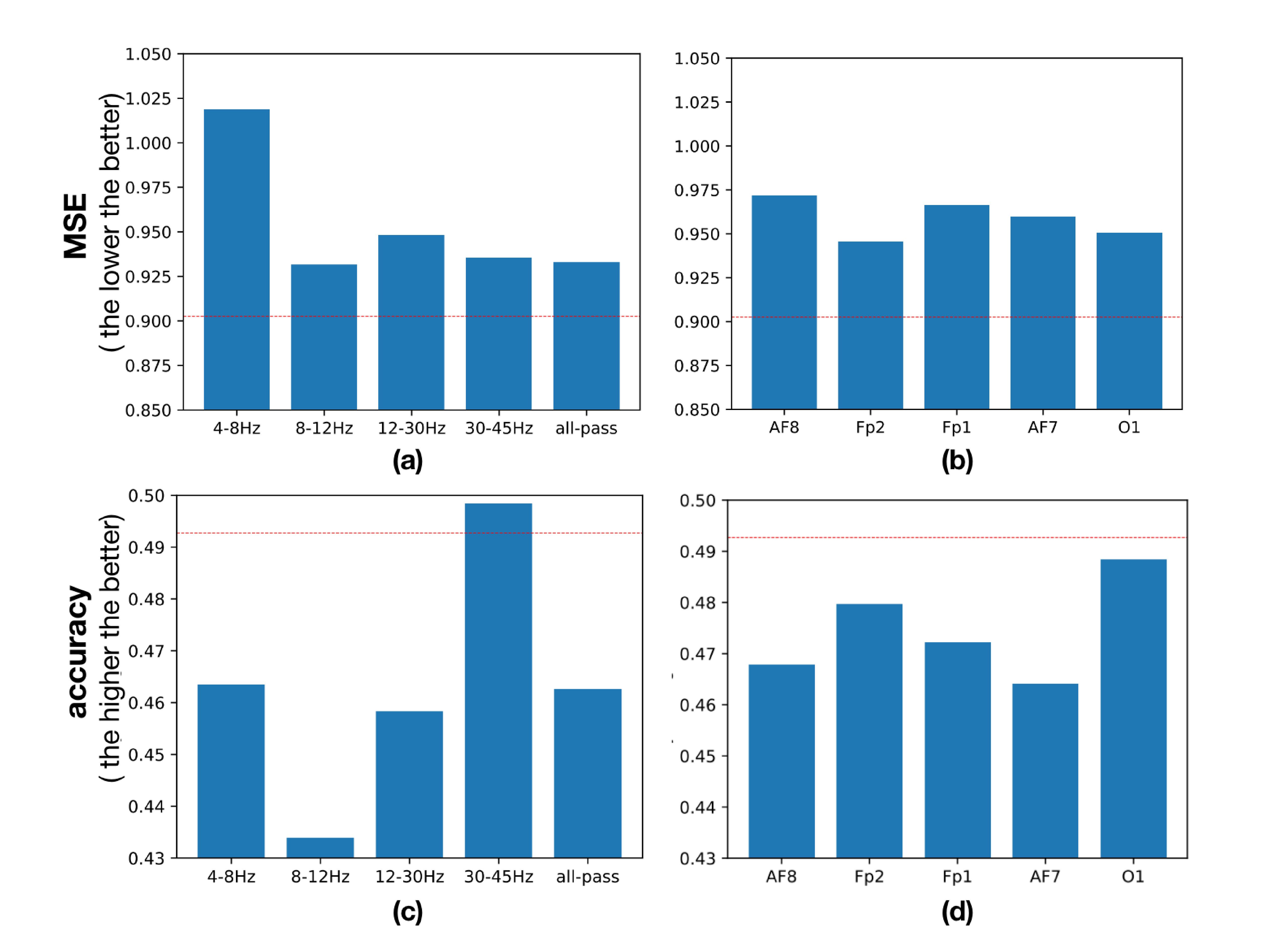}
\caption{The performance of the best model after removing each group of bands~(a and c) or each group of electrodes~(b and d) in the rating prediction~(a and b) and preference classification tasks~(c and d), respectively. Red dotted lines represent the performance with all the features.}
    \label{fig:ablation}
\end{figure*}

We conducted an ablation study to further analyse the usefulness of EEG features in the model. 
As shown in Figure~\ref{fig:ablation}, experiments are performed with removing each band or electrodes, that is, removing five features at a time. We held the experiments based on the best models with theirs settings and parameters for rating prediction and preference classification, respectively. 

As for the electrodes, whether in the rating prediction or the preference classification task, the first two models with the smallest performance loss are to remove Fp2 and O1, especially O1.  
As shown in Figure~\ref{fig:pref_corr}) in Section~\ref{sec:analysis}, only these two electrodes~(Fp2 and O1) did not have strongly correlated bands with preference. Besides, because the distance between O1 and the reference electrode O2 is the closest, and the position is symmetrical, it's explainable that O1 has no significant correlation with preference. Furthermore, AF8 and AF7 perform well for both tasks. 

As for the frequency bands, removing the 4-8 Hz band brings a great loss to the model performance. Besides, the 8-12 Hz band seems important to the preference classification model. 4-8 Hz and 8-12 Hz bands behave opposite over two tasks. We speculate this is due to differences in rating prediction and preference classification models. The activate function for preference classification is 'tanh', which adds nonlinearity to the model, while rating prediction is 'identity'. 
Besides, after removing the 30-45 Hz band in preference classification, the model performance is about the same as before~(p = 0.57 from the two-side t-test). The result demonstrates that the 30-45 Hz band has little effect on the model's performance in preference classification tasks. 

Two sum up, relatively uniform performance is present on the different electrodes for both tasks. Electrode AF8 and AF7 are important, while O1 and Fp2 perform poorly overall.
Difference frequency bands are useful to the model differently. Features performance changes due to the settings of the task. 4-8 Hz band is important to rating prediction while 8-12 Hz band is important to preference classification.

\subsubsection{Feature Importance Analysis}

The frequency bands and the electrodes are analyzed in the ablation study. Here we look at the specific features and the importance of PSD features in all features.

\begin{figure*}[ht]
    \centering
    \includegraphics[width=15cm]{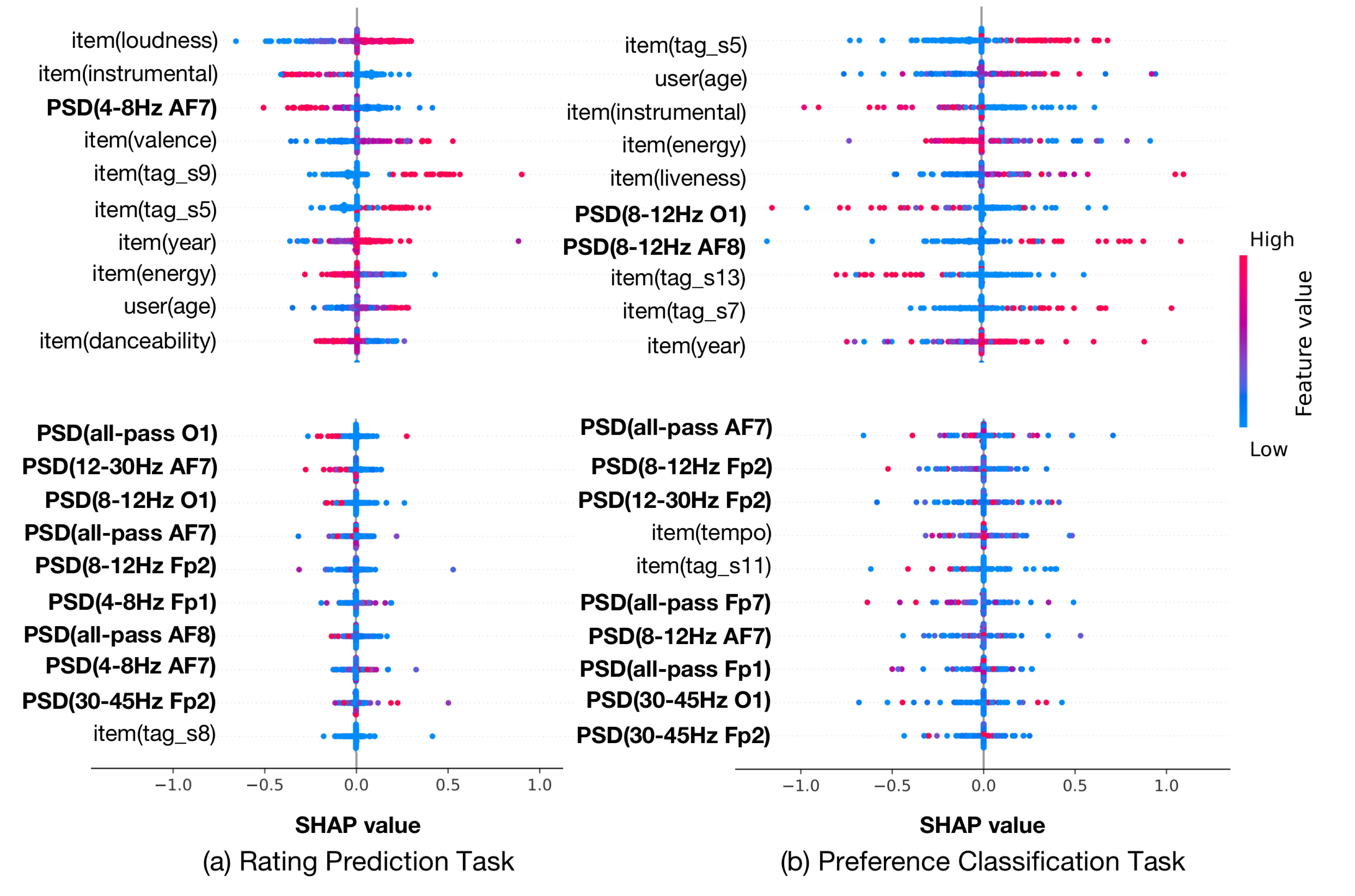}
\caption{SHAP values~(in terms of impact on model output) of top 10 and lowest 10 features for every sample in rating prediction task~(a) and preference classification task~(b), respectively. The features are sorted by their sum of SHAP magnitudes over fifty samples. The color represents the feature value (red high, blue low).}
    \label{fig:shap}
\end{figure*}

We randomly divide the train set and test set with the features of users, items, and EEG signals. Then we train the model~(MLP) for rating prediction and preference classification tasks on our best performance. Then, SHapley Additive exPlanations~\cite{shap}~((SHAP) is used to analyze the importance of each feature. We randomly select fifty samples and calculate the SHAP values of every feature for every sample. Then we sort them by the sum of SHAP magnitudes over all samples. The 
Figure~\ref{fig:shap} shows the top 10 and the lowest 10 features from fifty-seven features. SHAP values show the distribution of the impacts each feature has on the model output. 

As for the rating prediction task, among the top-10 important features, eight of ten are item features, one is the user's feature, and the other is the EEG  feature. It demonstrates that item features are irreplaceable. 
Besides, it reveals that a high PSD at AF7 of 4-8 Hz band lowers the predicted ratings obviously. It seems with the conclusion of the ablation study that electrode AF7 and the 4-8 Hz frequency band show importance. 
From the lowest-10 features, electrodes and bands vary. Features in the all-pass band are opposite to ratings, obviously. It illustrates the negative correlation between them and ratings, which we have analyzed in Section~\ref{sec:analysis}. 

As for the preference classification task, two PSD features are in the top 10. 
They are PSD from the 8-12 Hz band, showing the importance of the 8-12 Hz band in this task. The conclusion is the same as with the ablation study. 
Besides, it reveals the totally different contribution of PSD at O1 and AF8 of the 8-12 Hz band. High PSD in the former lowers the preference~(from positive to negative) and in the latter higher the preference~(from negative to positive).
Furthermore, the lowest-2 features are from the 30-45 Hz band, which is shown in the ablation study that it has little effect on the performance. 

Overall, some EEG features are more important than some user or item features, which indicates their significant supplement to the traditional recommendation using only user and item information.

\section{Discussion and Limitations}
\label{sec:discussion}

\subsection{Discussion}
Our work shows the potential to introduce EEG signals in personalized music recommendations with portable EEG devices. 
Here we discuss its contribution to personal music preferences, adoption of portable devices to record EEG signals, and privacy concerns. 

\subsubsection{EEG's Contribution to Personal Music Preferences}

Since EEG has recently demonstrated its value in various fields, some researchers are starting to apply it to the IR field. They use EEG signals to reveal relevance in information retrieval~\cite{liu2020challenges,chen2021search}. 
Besides, psychological research on affective computing with EEG illustrates the relationship between EEG and emotion. At the same time, there is an interactive relationship between music and emotions. Preferences and moods are related, so we use EEG signals to study people's preferences in the music scene, which is the basis of personalized recommender systems. To our knowledge, this is the first work that genuinely introduces EEG analysis into personalized music recommendations. 

In our work, we collected the data of 319 cases from 20 participants with EEG data when listening to music and feedback on preference and mood for listening to the music. PSD is extracted as the features of EEG signals. We analyze the relationship between EEG signals and preferences and the relationship between EEG signals and mood in Section~\ref{sec:analysis}. The results illustrate that EEG signals can distinguish moods well. Besides, EEG is associated with preference, which varies across electrodes and frequency bands.

And experiments in Section \ref{sec:Predction} illustrate that PSD extracted from portable EEG signals can help promote music preference prediction. There is significant improvement after introducing EEG to rating prediction and preference classification tasks via different typical models.
Further analysis demonstrates feature importance for rating prediction.
Therefore, promoting the prediction of music preference with the help of EEG signals is practical despite the cognitive gap between high-level music preference and portable EEG data. 

Moreover, our approach is not limited to preference prediction in music recommendations. Promising results with EEG signals also encourage applications of preference prediction in other scenarios. 

\subsubsection{Adoption of Portable Devices to Record EEG Signals}

In this work, we collected EEG signals with the dry-electrode portable device. 

Although EEG signals have been widely used in previous works for mood detection~(\cite{zheng2014emotion,soleymani2012multimodal}), wet-electrode with higher sampling frequency and much higher signal-to-noise ratio. So adoption of dry-electrode portable devices limits the utilization of EEG signals. However, it is of great value to attempt dry-electrode portable devices since they are closer to real-life applications in the foreseeable future than lab-based professional equipment such as wet-electrode devices.

It is a great challenge that the collected signals with dry-electrode EEG devices have low spatial resolution and high noise~(\cite{Lin2017foreheadEEG}). To solve the problem, we preprocessed the EEG signals by high-frequency filtering, normalization, and extracting typical features, reducing noise and getting clear information from EEG.
The encouraging results of our experiments show the possibility of adopting these portable EEG devices in research and application.

Meanwhile, we must point out that we believe the better results can be achieved with high-precision devices.
We are looking forward to more attempts in personal music preference with more accurate dry-electrode portable EEG devices, as the cost of these devices will fall with the development of technology.

\subsubsection{Privacy Concern in Personalized Recommendation}

Compared with conventional personalized recommendation methods, the opportunities of constructing personalized recommender systems come with risks for privacy challenges. 

On the one hand, no personally identifiable information is collected in the experiment. On the other hand, users must be informed about the content of the collected data, the method of collection, and the data usage. 
In our work, we explained the information collection and storage strategy with written informed consent and researchers face-to-face with each participant. 
They also agreed that their data would be used in the following experiments and made public after full anonymity. 

However, we must admit that our privacy protection strategy is not enough if the recommender system with EEG portable device is used on a large scale.
In the future, we will try to deal with the privacy issue from the perspective of data collection recommender models.

\subsection{Limitation}
Despite our best efforts, this work has several limitations. 

First of all, as our experiment was conducted limited to the stability of EEG apparatus, the scale of available data collection is only 319 cases from 20 participants.
The dataset sample size makes it impossible to conduct deeper and more effective models such as the state-of-the-art recommender system to better predict preference. Besides, limited by sample size, it's hard to analyze the relationship of PSD with preference and examine the usefulness of bracelet data more thoroughly.
Moreover, participants for the questionnaire survey and user study are mostly university students, so the demographic diversity is insufficient.
However, these limitations should be evaluated under the consideration that ours is the first study to introduce EEG signals in personal music preference and bring great potential for integration in the field of recommender systems. 

Besides, high data noise and rare electrodes give a hard-to-ignore impact, which is surely the disadvantage of dry-electrode portable devices. Although we try our best to reduce noise, it still brings unexplained differences over electrodes and frequency bands. Noise reduction signal decomposition may be a great problem in EEG study, but it is definitely a great challenge in our device. But based on the shared prospection of using EEG for daily life recommendation systems, the problems posed by dry electrodes need to be overcome. 

Lastly, the overall performance and ablation study shows limited promotion with EEG signals.
On the one hand, it is related to the low precision of the device discussed in the previous subsection.
On the other hand, we only extract PSD as the feature of EEG signals. 
We believe that more improvement can be achieved with the utilization of more EEG features and specific-designed methods for denoising in the future. 

\section{Conclusion and Future Work}
\label{sec:conclusion}

In this work, we introduce portable EEG data to personal music preferences. 
Then we conduct a user study with 20 participants to collect EEG signals with a dry-electrodes portable EEG device. EEG signals are collected in the process of listening to music, while feedback on mood and preference are also labeled. 
Based on the data collection, we preprocess the EEG signals with the extraction of PSD features. Then, we analyze the relationship between EEG signals and preference as well as mood. 
Experiments on rating prediction and preference classification demonstrate significant improvement with PSD features of EEG signals. Further analysis of the models shows the difference in importance of PSD over electrodes and frequency band. Theta ~(4-8 Hz)  band and alpha~(8-12 Hz) band shows the contribution to rating prediction and preference classification, respectively. 
The promising results of data analysis and experiments illustrate that introducing EEG signals to personal music preference in a recommendation scenario is possible. 
Our work provides a new direction for conducting music recommendations in daily life. Moreover, our approach is not limited to the music scenario. The EEG signals as explicit feedback~(in terms of preference) can be used in personalized recommendation tasks, which we leave as future work. 

\section*{Acknowledgments}
This work is supported by the Natural Science Foundation of China (Grant No. U21B2026) and Tsinghua University Guoqiang Research Institute.

We would like to thank all subjects who participated in the study and all partners in the THUIR lab for their support, assistance, and input throughout the research.

\bibliographystyle{ACM-Reference-Format}
\bibliography{reference}

\end{document}